\documentclass[11pt]{article}

\usepackage[utf8]{inputenc}
\usepackage[T1]{fontenc}
\usepackage[margin=1in]{geometry}
\usepackage{amsmath,amssymb}
\usepackage{booktabs}
\usepackage{caption}
\usepackage{array}
\usepackage{tabularx}
\usepackage{makecell}
\usepackage[authoryear,round]{natbib}
\usepackage[hidelinks]{hyperref}

\newcommand{\E}{\mathbb{E}}
\newcommand{\Prb}{\mathbb{P}}

\newcommand{\ind}{\mathbf{1}}

\title{Sample splitting for valid statistical inference in target trial emulations}
\author{Mats J. Stensrud\\[4pt]
\small Chair of Biostatistics, Institute of Mathematics,\\
\small \'Ecole polytechnique f\'ed\'erale de Lausanne (EPFL), Lausanne, Switzerland}
\date{}

\begin{document}

\maketitle

\begin{abstract}
\noindent Target trial emulation has improved comparative effectiveness research by making the causal question, assumptions, and analysis plan explicit. However, target trial protocols are usually developed iteratively. That is, after examining the data, investigators revise the protocol to reflect which target trials the observational data can realistically support. This iterative procedure is part of normal scientific practice, but it raises concerns about selective choices and invalid statistical inference. A simple procedure based on sample splitting addresses the concerns about valid inference. In the initial split, investigators can explore the data to define a target trial protocol. When these choices are made, the target trial protocol is implemented on the second split. Even though the investigators used the data to select the target trial protocol, the statistical inference has the usual coverage guarantees. The procedure intentionally mirrors how trialists move from pilot studies to a phase 3 trial. First, they use data from pilots and early-phase trials to decide on a final protocol. Then they implement this protocol and analyze a new set of data in a phase 3 trial.
\end{abstract}

% =====================================================================
\section*{Introduction}

When randomized trials are unavailable or infeasible, clinicians and policymakers often turn to observational data to inform decisions about the comparative effectiveness of interventions. Emulation of a target trial is a structured approach to causal inference from observational data \citep{hernan2016big}. Investigators specify the protocol of the hypothetical randomized trial that would answer their question of interest, often called the target trial, and then they use observational data to emulate that trial.

By making the hypothetical randomized trial explicit, we clarify the causal question by specifying all components of the target trial, including the eligibility criteria, treatment strategies, treatment assignment, follow-up, outcomes, causal contrasts, and analysis plan. Emulating a target trial also helps investigators avoid unnecessary errors, such as misaligned time zero, immortal time bias, and unclear eligibility criteria, which have plagued traditional observational studies \citep{hernan2016big,hernan2016specifying}. Target trial emulation does not, however, solve problems created by data limitations, such as unmeasured confounding, inadequate measurement of variables, or incomplete follow up. Instead, it improves the design of observational analyses so that they are more suitable to answer causal questions under explicit assumptions.

Applications of target trials are now appearing across many clinical areas. Recent reporting guidelines, such as the TARGET statement, further encourage investigators to explicitly describe the protocol of the target trial and how it was emulated \citep{hansford2025}. However, specifying an emulatable target trial is rarely a simple task that is done before the data are investigated. It is usually an iterative process in which investigators look at the data and then revise the target trial so that it is realistically emulatable \citep{hernan2026}. There are three main reasons for these iterations, and in practice they frequently occur together.

First, the available data constrain the set of causal estimands that can be formulated in terms of those data. Consider for example data collected during routine clinical interactions. The protocol of an emulatable target trial cannot include eligibility criteria that depend on tests only performed in research settings or treatments infrequently used in clinical practice. As investigators learn more about the available observational data, they adapt their original causal question until reaching a causal question that, arguably, can be formulated in terms of variables found in the data. During this iterative process, investigators choose the elements (e.g., eligibility criteria, treatment strategies, outcomes) of the trial protocols that they will attempt to emulate \citep{hernan2026}.

Second, the available data constrain the identifiability of the causal estimand. For example, suppose investigators want to emulate a target trial of cancer therapies and mortality in individuals with lung cancer. When inspecting the healthcare database, they find no information on certain clinical characteristics (e.g., performance status) that doctors use to recommend some cancer therapies. Because the absence of that information likely prevents successful adjustment for confounding, the investigators decide to adapt their question to one involving cancer therapies for which adequate data on confounders exist. That is, the treatment strategies of the target trial protocol are modified.

Third, the available data constrain what estimators can be implemented. For example, the available data often lead the investigators to make modelling choices. Consider investigators who end up adjusting for a subset of all available covariates after having confirmed that the point estimate is unaffected by discarding the other covariates. Then, the target trial protocol is decided after looking at the data. More broadly, investigators sometimes use the data to make choices in the statistical analysis, e.g., concerning the flexibility of estimators.

Inspecting the data before finalizing the protocol raises concerns about the validity of the statistical inferences. These concerns are different from those in trials. When the same observational data are used to iteratively refine the target trial protocol and to estimate the effect of the selected target trial, the conventional confidence intervals are not guaranteed to have their nominal coverage. In statistics, and in the quantitative sciences more broadly, it is well-known that using the data to choose estimands and models can deflate the coverage rates \citep[see, e.g.,][]{leamer1983,white2000,simmons2011}. This relates to known concerns about data snooping and selective reporting.

There is a simple approach that addresses those concerns: before investigators have had an opportunity to inspect the dataset, they can randomly split it into two samples. The ``specification sample'' is used to select an estimand and an estimator. The ``emulation sample'' is used to conduct the previously selected data analysis. The idea of using sample splitting has been around for decades. An early example is \citet{cox1975}, who suggested randomly dividing the data into two parts: one part for exploratory analysis and hypothesis formulation, and the other for formal significance testing, thereby overcoming difficulties that arise when the hypotheses tested are selected from the same data. In our context, splitting allows us to finalize the protocol and analysis using one part of the data and then conduct estimation and inference on the other part.

This procedure is related to, but different from, the conventional use of training and test samples in statistics and machine learning. The specification sample is not only used to choose an estimator, but also to define the causal estimand and the target trial protocol. The emulation sample is then used to estimate the selected estimand. Had the roles of the samples been exchanged, each specification sample could lead to a different estimand, so the resulting estimates would not necessarily target the same quantity.

% =====================================================================
\section*{When and why sample splitting works}

First, suppose that the protocol components were revised to align the causal question with what is available in routine clinical data. Such revisions do not necessarily invalidate conventional inference in the full sample. For example, if the protocol revisions are based only on which variables are present and how they are defined, rather than on the values they take or their association with the outcome, then inference can remain valid even when conducted on the whole dataset. Sample splitting nevertheless ensures that these protocol choices are made using the specification sample only, and that the effect estimate and confidence interval are computed in the emulation sample. The confidence intervals have their nominal coverage for the final estimand, regardless of the rationale for the protocol revisions. Sample splitting, however, does not ensure that the modified estimand is clinically ideal.

Second, suppose that protocol components were revised because the original estimand was not plausibly identifiable with the available covariates. In the lung cancer example, suppose that performance status is not recorded even though it affects treatment choices, so investigators switch to a comparison for which confounding control, and thus identification, is more plausible. This revision, based only on the fact that a confounder is not recorded, does not necessarily invalidate conventional inference in the entire sample. However, such revisions are problematic when they depend on features of the observed covariates, such as learning that a confounder is frequently missing in certain subgroups. Sample splitting solves the inference problem, because the estimate and confidence interval are computed in a separate sample.

Third, suppose that protocol components were changed because of estimation feasibility or statistical precision. Then, conventional statistical inference is usually invalid because the protocol is selected based on the same data used for estimation, with only special cases as exceptions \citep[][gave one example with propensity scores]{rosenbaum1983}. Suppose analysis choices were revised in response to estimator stability and modelling considerations, such as convergence problems, selection among functional forms or link functions, or choices about covariate adjustment. These modelling choices generally invalidate conventional confidence intervals computed from the entire dataset. Sample splitting ensures that such modifications are treated as fixed before analyzing the emulation sample.

\section*{Estimand picking}

A potential concern in target trial emulation is that investigators examine several target trial protocols and report only those analyses that yield the most promising results or the largest estimated effects. Many investigators know that such practices are problematic and avoid them, but the risk of such estimand picking is still present.

In the lung cancer example, suppose that the investigators explore different treatment strategies, eligibility criteria, or outcomes and then select a protocol that appear to have large benefit in the exploratory analysis. This selection is problematic for the interpretation of the classical confidence intervals, as the selection depends on data on outcomes and anticipated effects. However, sample splitting makes the reported effect estimate and confidence interval come from data that were not used to select the protocol. Thus, the final confidence interval retains its nominal coverage. This is true even if the investigators use treatment--outcome associations and preliminary effect estimates to choose their estimand in the specification sample. This type of ``estimand picking'' allows the investigator to shift attention toward outcomes where effects are more prominent and away from outcomes where effects are small or null, regardless of sample splitting. However, such selective reporting is not unique to emulated target trials, but likely appears in RCTs for drug development too: based on laboratory studies and early phase trials, drug developers intentionally create inclusion criteria and primary outcomes for phase 3 trials such that effects are expected to be larger or events more frequent. For example, many cancer trials use cancer-specific mortality as the primary outcome because it gives more power than all-cause mortality. Relatedly, outcomes such as quality of life are difficult to measure compared to hard outcomes like mortality. The softer outcomes are therefore rarely primary endpoints. Whenever there is room for selective estimand picking, the investigators should describe the choices made in the initial split.

% =====================================================================
\section*{Inference guarantees and why transparency is not enough}

There exists work on adaptive prespecification of study protocols, which gives guidance for stating the initial target trial protocol, describing foreseeable protocol changes, and documenting the reasons for going from the initial protocol to the final target trial emulation \citep[see, for example,][]{weckstein2025,muntner2024}. Sample splitting does not replace this work, but can be regarded as complementary. Suppose that the investigator prespecifies that the specification sample will be used to examine whether the outcome is recorded, whether treatment strategies are sufficiently common, whether key confounders are measured, whether follow-up is adequate, and whether positivity is plausible in the intended target population. If the specification sample suggests that the initial target trial is hard to emulate, then the protocol can be changed and these changes can be reported. The final protocol can be implemented in the emulation sample to ensure nominal statistical guarantees.

An alternative is to avoid sample splitting, register the initial target trial protocol, report every amendment, and explain why the final emulated trial differs from the initial one. This strategy ensures some transparency, but it does not solve the statistical problem. Suppose that the investigator inspects the data, changes the eligibility criteria, treatment strategies, outcome, follow-up, causal contrast, or analysis plan, and then estimates the effect using the same data. The conventional confidence interval does not necessarily have its usual coverage, even when the amendment was motivated by feasibility rather than by a search for a large effect.

% =====================================================================
\section*{Limitations of sample splitting}

A downside of sample splitting is that only part of the data is used for the final analysis. This is particularly problematic in studies with a small number of individuals. There is no simple solution to improve power and at the same time ensure valid inference. Most design choices about eligibility, treatment strategies, follow-up, and outcomes invalidate classical inference in the whole dataset. This raises a question about the optimal splitting fraction. There is no single answer to this question \citep[see, e.g.,][for a particular set-up]{yin2026}, as it depends on the true data generating process and the data available. However, using 10--20\% of the sample is likely enough to make protocol decisions in studies with more than 1000 subjects. This rough consideration also agrees with recommendation in related sample splitting settings \citep{heller2009}. Also, cross-fitting, which repeatedly swaps the roles of the data splits so that each observation is used for model fitting in some folds and for estimation in others, is not immediately suitable in our context. This is because the first split is used to choose the final protocol, not just to fit statistical models. However, some prespecified changes might be handled with methods for adaptive trials that use the whole dataset, but those methods are rarely tailored to decisions about target trial protocols and require specialized inference. There are also more sophisticated approaches that, in principle, allow reuse of the second split data in specific ways, for example, based on ideas from differential privacy \citep{dwork2015}. Such methods, however, are harder to implement and are less accessible.

We also face another problem that rarely happens in randomized trials, except perhaps for recent platform trials. When emulating a target trial, the investigators have often worked with parts of the same data in earlier studies. These earlier studies could, for example, be emulations of other target trials with different treatments but similar outcomes. This opens the possibility that the investigators use prior knowledge about the outcomes when designing the new protocol. Sample splitting does not fully protect against undercoverage in this setting, because the second split cannot be regarded as entirely new or unseen. Still, we believe that any undercoverage from prior familiarity is smaller than the undercoverage from using the same data twice in a single analysis.

Furthermore, the sample splitting strategy is only relevant when changes to the protocol are scientifically or practically acceptable. Consider an investigator who perfectly specifies a target trial protocol before looking at the study data. When preparing the analysis, they realize that the ideal outcome is not recorded, an important confounder is absent, one treatment strategy is rarely used in the data, or the intended population has severe positivity problems. One option is to conclude that the available data cannot answer the original practical question, and no related question is of interest to them. Therefore, they give up the analysis. Another option is to construct a target trial that answers a question close to the original practical problem, for example by restricting the analysis to a narrower target population or by evaluating another related outcome.

% =====================================================================
\section*{Numerical examples}

To illustrate when sample splitting is useful, we will consider simple examples with an investigator who aims to compare two treatments for patients with metastatic non-small cell lung cancer. Suppose, first, that quality of life was the primary outcome of interest in an observational study. When inspecting the data, the investigators found that quality of life measurements were not collected in a form that could be analyzed. However, they had access to all-cause mortality data, which was the only prespecified fallback outcome. They changed the estimand to all-cause mortality. In this case, the final estimand was not selected from competition among several analyses based on empirical results, but because of lacking measurements. Therefore, the ordinary inference results are not necessarily invalid (see Table~\ref{tab:main}, row 1, and also Example~1 in the supplementary material for a concrete calculation).

However, a full sample analysis is problematic when the final protocol is chosen because it looks more feasible. To illustrate the problem, we consider additional settings.

Suppose that an investigator has decided to study the average treatment effect before looking at the data, and has found several valid specifications of the outcome regression model, for example with different functional forms or different sets of adjustment covariates. In the data analysis, they fit all candidate specifications and report the one with the smallest estimated standard error. Even if every candidate is correctly specified and equally precise, this selection reduces coverage. With 16 candidate specifications, coverage drops from 95\% to about 93.4\% with 200 patients per treatment group (Table~\ref{tab:main}, row 2). The undercoverage increases with the number of candidates and decreases with the sample size (Supplementary Table~\ref{tab:smallestse}).

Suppose now that the investigator instead starts by implementing a flexible outcome regression model, which is correctly parameterized. However, this model is only reported if there are no convergence or stability issues. Otherwise, the investigator will use a simpler regression, which also is correctly parameterized. To mimic this setting numerically, consider a strategy where the flexible model is used when the standard deviation is smaller than a prespecified constant. In the analyses that retain the flexible model, the reported confidence intervals undercover (Table~\ref{tab:main}, row 3).

Consider an investigator who also studies the average treatment effect. They fit two outcome regression models, one model with many covariates and interaction terms (the full model), and another model with no interaction term. They have beforehand decided to report the reduced model as their main result if the point estimate is close to that from the larger model. This strategy also leads to undercoverage, even if the reduced model is correctly specified (Table~\ref{tab:main}, row 4) and also when it has bias (Table~\ref{tab:main}, row 5).

\begin{table}[!htbp]
\centering
\caption{Examples illustrating coverage (nominal coverage 95\%)}
\label{tab:main}
\small
\begin{tabularx}{\textwidth}{@{}>{\raggedright\arraybackslash}p{0.30\textwidth} >{\raggedright\arraybackslash}X >{\centering\arraybackslash}p{0.11\textwidth}@{}}
\toprule
Analytic strategy & Selection rule & Coverage \\
\midrule
A prespecified backup outcome & Use mortality when quality of life was not recorded in an analyzable form & 95.0\% \\
Several equally valid model specifications & Report the specification with the smallest estimated standard error & 93.4\% \\
Flexible model as first choice & Use the flexible regression model if the standard deviation is smaller than a prespecified value & 93.4\%$^{*}$ \\
Choose a small model if it agrees with a larger, flexible model & Use the reduced model if its estimate is similar to the one from the full model & 92.8\% \\
Choose between a full model and a biased reduced model & Use the reduced model if its estimate is similar to the one from the full model & 90.1\% \\
\bottomrule
\end{tabularx}
\par\smallskip
\begin{minipage}{\textwidth}
\footnotesize $^{*}$Coverage of the intervals reported when the flexible model is retained, with 200 patients per treatment group.
\end{minipage}
\end{table}

% =====================================================================
\section*{On the magnitude of distortion}

One way of thinking about post-selection distortion is to imagine repeating the entire analysis in many new samples from the same target population. If we would select the same protocol in every sample, then this data-driven selection is essentially non-random. Thus, conventional statistical inference for this protocol is valid, in the sense that it gives the usual statistical guarantees. In particular, a 95\% confidence interval would have nominal coverage of the selected estimand. Suppose now that one protocol is selected with at least probability $1-\varepsilon$, and the investigator uses a conventional inference procedure that is only guaranteed to be valid when the protocol is fixed. Then, the distortion of the coverage is at most $\varepsilon$. Thus, when one protocol is selected with high probability, the distortion of coverage due to selection must be small. For example, if the same protocol is chosen in at least 99\% of samples, the coverage is at least 94\%, and likely better in many practical settings.

Suppose that an investigator chooses among several candidate outcomes by reporting the one with the largest estimated average treatment effect. Suppose also that the true effect on one of the outcomes is several standard errors larger than the effects on all the other outcomes. Then, the outcome with largest effect will be selected in almost every random sample from the target population. In that case, the conventional inference will give close to nominal coverage. Alternatively, suppose that the true effect on many of the candidate outcomes is similar. Then, the selected outcome is sensitive to random sampling variation and conventional confidence intervals can undercover. This consideration also helps understand limitations described in the section ``Limitations of sample splitting''. For example, suppose that an investigator looks at the data and selects a primary outcome because the treatment effect estimate was large. Suppose also that this outcome would be selected in almost every sample, so that conventional inference for that outcome has nearly nominal coverage. Then, sample splitting does not guarantee that the estimand is practically interesting and will lead to estimators with lower precision.

More broadly, it would be interesting to better understand the exact undercoverage of full sample analyses, as there are likely other settings where the undercoverage is negligible. The undercoverage can also be small in settings with substantial random selection, for example when selection is based on certain procedures that are valid conditional on the selection. This future work can draw on existing results on post-selection inference \citep[see, e.g.,][]{kuchibhotla2022}.

% =====================================================================
\section*{Conclusion}

Target trial emulations benefit from a clear separation between the protocol choices and the statistical analysis. This can be done by using one data split to define the protocol and a second split to estimate causal effects. The splitting strategy reflects how trialists use pilot and early-phase studies to finalize a protocol before implementing a phase 3 trial. Design choices based on the initial split can make identification assumptions more plausible and improve precision. The clear advantage of this procedure is that it lets investigators construct the target trial protocol based on data while still applying classical methods for statistical inference.

% =====================================================================
\section*{Acknowledgements}

I would like to thank Marco Piccininni and Miguel A. Hern\'an for their helpful comments and suggestions. I used ChatGPT versions 5.5 and 5.6 and Claude Fable 5.1 to help with the language and the structure of the manuscript. I also used ChatGPT 5.5 and Claude Fable 5.1 to help constructing the numerical examples. I have verified and edited the suggestions given by the large language models, and I take full responsibility for the content in the manuscript.

% =====================================================================

% =====================================================================
% SUPPLEMENTARY MATERIAL
% =====================================================================
\clearpage
\setcounter{table}{0}
\renewcommand{\thetable}{S\arabic{table}}

\begin{center}
{\Large\bfseries Supplementary material}\\[4pt]
{\large Sample splitting for valid statistical inference in target trial emulations}
\end{center}

\section*{Examples}

The examples below elaborate on the sections ``Numerical examples'' and ``On the magnitude of distortion'' in the main text. In all of these examples, suppose that the investigator wants to compare two treatments for patients with metastatic non-small cell lung cancer. Throughout, \(\Phi\) and \(\phi\) denote the standard normal distribution and density functions, and \(t_{0.975,\nu}\) denotes the 97.5th percentile of the \(t\) distribution with \(\nu\) degrees of freedom.

\subsection*{Example 1. A single alternative outcome}

Suppose that the intended primary outcome is quality of life at one year. Before any outcome analysis is performed, the investigators inspect the documentation of the database and realize that quality of life was not collected in a form that can be analyzed. The protocol had specified one alternative outcome, one-year mortality, and the investigators therefore decide to use this as their primary outcome.

The choice is not determined by the observed treatment effects, standard errors, event counts, convergence or model behavior. Thus, the same alternative outcome would be chosen in repeated samples from this database, and the choice does not distort the coverage of the confidence interval.

To be explicit, suppose the mortality effect estimator is normally distributed,
\[
\widehat\psi_M \sim N(\psi_M,1).
\]
The usual 95\% confidence interval is
\[
\widehat\psi_M \pm 1.96.
\]
Coverage for the revised estimand \(\psi_M\) is
\[
\Prb\{\lvert\widehat\psi_M-\psi_M\rvert\leq 1.96\}
=
\Prb\{\lvert Z\rvert\leq 1.96\}
\approx
0.95,
\]
where \(Z\sim N(0,1)\).

\subsection*{A common model for Examples 2 and 3}

Suppose that the estimand is the average treatment effect, \(\psi=\E\{Y(1)-Y(0)\}\), and that an outcome regression is fitted with \(m\) patients in each treatment group. Suppose that the effect estimator \(\widehat\psi\) and the estimated residual standard deviation \(S\) satisfy
\[
\widehat\psi-\psi \sim N\!\left(0,\frac{2\sigma^2}{m}\right),
\qquad
\frac{\nu S^2}{\sigma^2}\sim\chi^2_{\nu},
\qquad
\nu = 2m-2,
\]
with \(\widehat\psi\) independent of \(S\), as in a normal linear model with homoscedastic errors. For a model that is fixed in advance, the usual 95\% \(t\) interval
\[
\widehat\psi \pm t_{0.975,\nu}\,S\sqrt{\frac{2}{m}}
\]
has 95\% coverage. Conditional on \(S=s\), the interval covers \(\psi\) with probability
\[
\Prb\left\{\left|\widehat\psi-\psi\right|\leq t_{0.975,\nu}\,s\sqrt{\frac{2}{m}}\ \middle|\ S=s\right\}
=
2\Phi\!\left(\frac{t_{0.975,\nu}\,s}{\sigma}\right)-1,
\]
which is increasing in \(s\). Selection rules that favor small values of \(S\) therefore lead to reported intervals that are too narrow. In the numerical results we set \(\sigma=10\). None of the coverage probabilities in Examples 2 and 3 depend on the value of \(\psi\).

\subsection*{Example 2. Choosing among equally valid model specifications by the smallest estimated standard error}

Investigators often have several defensible specifications of the outcome model, for example different functional forms, link functions, or covariate sets, all of which are valid for the same estimand. Suppose that the investigator fits \(K\) such specifications and reports the one with the smallest estimated standard error.

To be explicit, suppose that for each candidate specification \(j=1,\ldots,K\), the estimator and its residual standard deviation follow
\[
\widehat\psi_j-\psi \sim N\!\left(0,\frac{2\sigma^2}{m}\right),
\qquad
\frac{\nu S_j^2}{\sigma^2}\sim\chi^2_{\nu},
\]
so that all candidates are unbiased and equally precise, and assume that the vector of effect estimators \((\widehat\psi_1,\ldots,\widehat\psi_K)\) is independent of the vector of standard deviations \((S_1,\ldots,S_K)\). To simplify the calculations, assume also that \(S_1,\ldots,S_K\) are independent (This is implausible, because analyses of the same data will have positively correlated standard errors, and positive correlation makes the smallest of the \(S_j\) less extreme, so the coverage violation will typically be smaller than the one reported below, but the phenomenon is the same).

Define
\[
\widehat\jmath=\arg\min_{1\leq j\leq K} S_j,
\qquad
S_{\min}=S_{\widehat\jmath}.
\]
Conditional on \((S_1,\ldots,S_K)\), the selected index \(\widehat\jmath\) is fixed, and because the effect estimators are independent of the standard deviations, \(\widehat\psi_{\widehat\jmath}-\psi\) still has variance \(2\sigma^2/m\). Therefore, conditional on \(S_{\min}=s\), the selected interval covers with probability \(2\Phi(t_{0.975,\nu}s/\sigma)-1\), and the coverage of the conventional procedure that ignores the selection is
\[
\E\left\{2\Phi\!\left(\frac{t_{0.975,\nu}\,S_{\min}}{\sigma}\right)-1\right\}.
\]
Because \(S_{\min}\) is selected to be unusually small, this expectation is below 0.95. As \(m\) increases, each \(S_j\) concentrates around \(\sigma\), so the undercoverage becomes smaller for a fixed number of candidate models \(K\). Table~\ref{tab:smallestse} evaluates the expectation from the distribution of \(S_{\min}\).

With four candidate models, choosing the analysis with the smallest estimated standard error reduces coverage from 95\% to about 92.1\% when \(m=20\) and to about 94.1\% when \(m=200\). With 16 candidate models, the corresponding coverages are about 89.1\% and 93.4\%. The coverage problem becomes more severe when more candidate analyses are considered, and less severe in larger samples.

\begin{table}[!htbp]
\centering
\caption{Choosing, among \(K\) equally valid specifications, the analysis with the smallest estimated standard error (\(\sigma=10\); nominal coverage 95\%)}
\label{tab:smallestse}
\small
\begin{tabular}{@{}c cc cc@{}}
\toprule
& \multicolumn{2}{c}{\(m=20\) per treatment group} & \multicolumn{2}{c}{\(m=200\) per treatment group} \\
\cmidrule(lr){2-3}\cmidrule(lr){4-5}
\makecell{Candidate\\specifications \(K\)} & \makecell{Expected\\selected \(S\)} & Coverage & \makecell{Expected\\selected \(S\)} & Coverage \\
\midrule
1  & 9.93 & 95.0\% & 9.99 & 95.0\% \\
2  & 9.29 & 93.5\% & 9.79 & 94.5\% \\
4  & 8.77 & 92.1\% & 9.63 & 94.1\% \\
8  & 8.34 & 90.6\% & 9.49 & 93.8\% \\
16 & 7.97 & 89.1\% & 9.37 & 93.4\% \\
32 & 7.65 & 87.7\% & 9.27 & 93.1\% \\
\bottomrule
\end{tabular}
\end{table}

\subsection*{Example 3. Model convergence and stability diagnostics}

Investigators often change the analysis after a flexible estimator fails to converge, gives unstable predictions, or produces unsatisfactory diagnostics. To illustrate the consequences, consider the common model with a single specification, and suppose that a flexible outcome regression is reported only when a diagnostic check is favorable; to be concrete, let the check be the event \(S\leq c\), where \(S\) is the estimated residual standard deviation. Otherwise, a simpler regression is used. The intervals that are reported when the flexible model is retained have conditional coverage
\[
\E\left[
2\Phi\!\left(\frac{t_{0.975,\nu}S}{\sigma}\right)-1
\,\middle|\,
S\leq c
\right],
\]
which is below 0.95 because conditioning on \(S\leq c\) favors small values of \(S\). Table~\ref{tab:stability} gives the probability that the flexible model is retained and the conditional coverage of the reported interval.

This conditional coverage describes the intervals given that the flexible model was retained. The global coverage of this procedure also depends on the coverage when \(S>c\). If the simpler regression is valid and its interval does not depend on \(S\), the coverage is
\[
\E\left[
\left\{2\Phi\!\left(\frac{t_{0.975,\nu}S}{\sigma}\right)-1\right\}\ind\{S\leq c\}
\right]
+0.95\,\Prb(S>c),
\]
which is given in the last column of Table~\ref{tab:stability}. In the other extreme case, in which the simpler regression yields the same interval as the flexible model, the coverage of the procedure is exactly 95\%. Thus, the selection rule distorts the coverage of the intervals that are reported under the flexible model, whereas the procedure as a whole remains approximately valid.

\begin{table}[!htbp]
\centering
\caption{Coverage under a selection rule based on diagnostics (\(\sigma=10\); nominal coverage 95\%)}
\label{tab:stability}
\small
\begin{tabular}{@{}cc ccc@{}}
\toprule
\makecell{\(m\) per\\group} & \makecell{Selection\\rule} & \makecell{Probability flexible\\model retained} & \makecell{Conditional coverage\\if retained} & \makecell{Coverage of the\\procedure\(^{\dagger}\)} \\
\midrule
20  & \(S\leq 10.5\) & 69.4\%    & 93.8\% & 94.2\% \\
20  & \(S\leq 9.5\)  & 35.9\%    & 92.1\% & 94.0\% \\
20  & \(S\leq 9.0\)  & 20.9\%    & 90.9\% & 94.1\% \\
20  & \(S\leq 8.5\)  & 10.3\%    & 89.3\% & 94.4\% \\
\addlinespace
200 & \(S\leq 10.5\) & 92.3\%    & 94.9\% & 94.9\% \\
200 & \(S\leq 9.5\)  & 8.1\%     & 93.4\% & 94.9\% \\
200 & \(S\leq 9.0\)  & 0.22\%    & 92.0\% & 95.0\% \\
200 & \(S\leq 8.5\)  & 0.0008\%  & 90.2\% & 95.0\% \\
\bottomrule
\end{tabular}
\par\smallskip
\begin{minipage}{0.92\textwidth}
\footnotesize \(^{\dagger}\)Assuming that the simpler regression used when \(S>c\) yields a valid 95\% interval that does not depend on \(S\).
\end{minipage}
\end{table}

\subsection*{Example 4. Choosing a simpler model after checking that the estimate is unchanged}

Some investigators remove covariates or simplify a parameterization after checking that the point estimate is insensitive to these changes. The following example illustrates that this practice can be problematic.

Suppose that the full model and the reduced model are both valid for the same estimand. At a particular sample size, let
\[
\widehat\psi_F \sim N(\psi,1^2),
\qquad
\widehat\psi_R \sim N(\psi,0.5^2),
\]
with correlation 0.5, so the reduced model is more precise. Suppose that the investigator reports the reduced model if its estimate is close to the full-model estimate,
\[
\left|\widehat\psi_R-\widehat\psi_F\right|\leq 0.5,
\]
and otherwise reports the full model. If either model were fixed in advance, its usual 95\% interval would have 95\% coverage.

Because
\[
\widehat\psi_R-\widehat\psi_F
\sim
N(0,0.75),
\]
the reduced model is selected with probability
\[
\Prb\left(\left|\widehat\psi_R-\widehat\psi_F\right|\leq 0.5\right)
=
2\Phi\left(\frac{0.5}{\sqrt{0.75}}\right)-1
\approx
0.436.
\]
Without loss of generality set the true effect to be \(\psi=0\). The coverage of the reported interval is
\begin{align*}
&\Prb\left(
|\widehat\psi_R|\leq 1.96\times 0.5,
\ |\widehat\psi_R-\widehat\psi_F|\leq 0.5
\right)\\
&\quad +
\Prb\left(
|\widehat\psi_F|\leq 1.96,
\ |\widehat\psi_R-\widehat\psi_F|>0.5
\right)
\approx
0.9284.
\end{align*}
Thus the selection procedure has about 92.8\% coverage, but both candidate intervals have 95\% coverage without selection.

The problem can increase if the reduced model is biased. For example, suppose that
\[
\widehat\psi_R \sim N(\psi+0.4,0.5^2),
\]
and that the full model remains unbiased and the covariance structure is unchanged. Now, the reduced model is selected with probability approximately 0.40. The bias of the reported estimator is approximately 0.017, and the coverage of its nominal 95\% interval is approximately 0.90.

\begin{table}[!htbp]
\centering
\caption{Choosing a reduced model after checking that its estimate is close to the full-model estimate (nominal coverage 95\%)}
\label{tab:reduced}
\small
\begin{tabularx}{\textwidth}{@{}>{\raggedright\arraybackslash}X *{3}{>{\centering\arraybackslash}p{0.16\textwidth}}@{}}
\toprule
Setting
& \makecell{Reduced model\\selected}
& \makecell{Bias of\\reported\\estimator}
& \makecell{95\%\\coverage} \\
\midrule
Both models valid & 43.6\% & 0.00 & 92.8\% \\
Reduced model biased by 0.4 & 39.7\% & 0.02 & 90.1\% \\
\bottomrule
\end{tabularx}
\end{table}

\subsection*{Example 5. When selection is nearly deterministic}

Let \(J\) denote the selected protocol after data inspection, and let \(C_j\) denote the event that the nominal 95\% confidence interval for protocol \(j\) covers \(\psi_j\). Suppose that a particular protocol, \(j^\star\), is selected with probability at least \(1-\varepsilon\), and that each interval has exact 95\% coverage without selection. Because the event that the selected protocol \(J\) covers, \(C_J\), and the event that the fixed protocol \(j^{\star}\) covers, \(C_{j^\star}\), are identical whenever \(J=j^\star\), we have that
\[
\left|
\Prb(C_J)-0.95
\right|
=
\left|
\Prb(C_J)-\Prb(C_{j^\star})
\right|
\leq
\Prb(J\neq j^\star)
\leq
\varepsilon.
\]
Thus, if the same protocol is selected with probability close to one, the coverage distortion caused by selection has to be small, however the selection is made.

To be explicit, suppose there are four candidate outcomes. The effect estimators are independent and satisfy
\[
\widehat\psi_j\sim N(\psi_j,1),
\qquad j=1,\ldots,4,
\]
and the investigator reports the outcome with the largest estimated effect. Let \(\psi_1=\Delta\) and \(\psi_2=\psi_3=\psi_4=0\). Each interval \(\widehat\psi_j\pm1.96\) has 95\% coverage. The probability that outcome 1 is selected is
\[
\Prb(J=1)
=
\int_{-\infty}^{\infty}
\phi(z)\Phi(z+\Delta)^3\,dz.
\]
The coverage of the interval for the selected outcome is
\[
\int_{-1.96}^{1.96}
\phi(z)\Phi(z+\Delta)^3\,dz
+
3\int_{-1.96}^{1.96}
\phi(z)\Phi(z)^2\Phi(z-\Delta)\,dz.
\]

\begin{table}[!htbp]
\centering
\caption{Coverage, which depends on the probability of selecting a particular outcome (nominal coverage 95\%)}
\label{tab:deterministic}
\small
\begin{tabular}{@{}ccc@{}}
\toprule
\(\Delta\) & \(\Prb(J=1)\) & Coverage of the selected confidence interval \\
\midrule
0 & 25.0\% & 90.4\% \\
1 & 55.2\% & 90.9\% \\
2 & 82.3\% & 92.8\% \\
3 & 95.6\% & 94.5\% \\
4 & 99.3\% & 95.0\% \\
\bottomrule
\end{tabular}
\end{table}

The results show that, when the candidate outcomes have similar true effects, the selected outcome varies substantially across samples and the nominal interval can undercover. When one effect is several standard errors larger than the alternatives, the same outcome is selected in almost every sample and coverage approaches 95\%. The bound \(\varepsilon\) is often loose: for \(\Delta=3\), \(\varepsilon=0.044\), whereas the actual distortion is only 0.5 percentage points.

\end{document}